\documentclass[aps,prl,twocolumn,letter,longbibliography,floatfix]{revtex4-1}

\usepackage{graphicx}
\usepackage{dcolumn}
\usepackage{bm}

\usepackage{float}
\usepackage{amsmath}
\usepackage{amsfonts}
\usepackage{color}
\usepackage[normalem]{ulem}
\usepackage[english]{babel}
\DeclareUnicodeCharacter{2009}{\,}

\newcommand{\be}{\begin{equation}}
\newcommand{\ee}{\end{equation}}

\newcommand{\jamesemph}[1]
{\textcolor{blue}{#1}}

\definecolor{mygray}{gray}{0.4}

\newcommand{\cn}[1]
{\jamesemph{[Citation Needed!]}}

\begin{document}

\title{Predicting Defects in Soft Sphere Packings Near Jamming Using The Force Network Ensemble}
\author{James D. Sartor$^*$, Eric I. Corwin$^*$}
\affiliation{$^*$Department of Physics and Materials Science Institute, University of Oregon, Eugene, Oregon 97403, USA }

\date{\today}

\begin{abstract}
Amorphous systems of soft particles above jamming have more contacts than are needed to achieve mechanical equilibrium. The force network of a granular system with a fixed contact network is thus underdetermined and can be characterized as a random instantiation within the space of the force network ensemble. In this work, we show that defect contacts which are not necessary for stability of the system can be uniquely identified by examining the boundaries of this space of allowed force networks. We further show that in the near jamming limit, this identification is nearly always correct and the defect contacts are broken under decompression of the system.
\end{abstract}

\maketitle

From crafting swords and arrowheads in antiquity to perusing katanas at the mall today, choosing the available materials with highest strength has always been of critical concern. It is the weak points, and modes of failure, that determine the strength of a material. In polycrystalline materials these weak points arise from defects in the crystal structure \cite{taylor_mechanism_1934}. There are many forms of crystalline defects, and methods such as annealing, tempering, and quenching are designed to manipulate these defects, making a material stronger.
Early approaches to amorphous systems attempted to model them as highly defective crystalline systems, but such models fail to capture emergent phenomena \cite{goodrich_solids_2014}. Thus amorphous systems must be treated in their own right as a qualitatively separate system, and consequently, there exists no obvious definition of a defect. Amorphous materials do however have structural defects, termed ``soft spots'', which are locations in which rearrangements are more likely to occur under quasistatic shear. These were first explored via analysis of the low-frequncy quasilocalized vibrational modes \cite{manning_vibrational_2011} and have been more recently identified by using machine learning analysis on the local structure \cite{cubuk_identifying_2015, cubuk_structural_2016, ding_soft_2014, wijtmans_disentangling_2017,schoenholz_structural_2016,schoenholz_understanding_2014,richard_predicting_2020,rocks_learning-based_2021,ridout_correlation_2020}.
While this has been highly effective at identifying sites of rearrangements in both athermal and thermal jammed systems under shear, it has not yet been applied to systems under decompression, another common failure mode of materials. More importantly, while softness is correlated with structural quantities such as local potential energy and coordination number, these structural properties are not good predictors of rearrangements on their own. Thus softness, while useful as a heuristic, lacks analytic clarity. Additionally, while softness is an excellent predictor of instabilities, it does not predict stable contact network changes (i.e. contact changes which do not result in rearrangements), which comprise the majority of contact network changes \cite{morse_differences_2020,tuckman_contact_2020}. In this work we demonstrate a method for identifying defective contacts under decompression asymptotically close to the jamming/unjamming transition. We use the geometry of the force network structure to show that in the very near-jamming limit, there exists only a small and precisely identifiable number of contacts at which any contact network change can possibly occur.

\begin{figure}[t]
\includegraphics[width=\columnwidth, trim=99 240 136 255, clip]{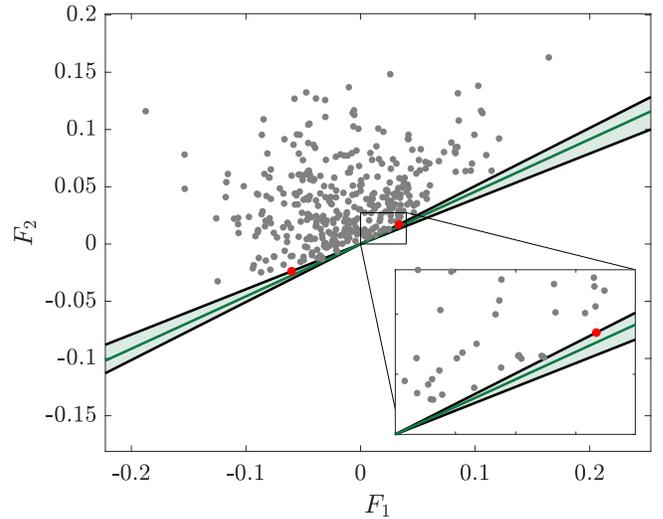}
\caption{A scatter plot of the loads on each contact in the two states of self stress $F_1$ and $F_2$ for a typical system ($N=128$) at 2SSS. The distance from the axes represents the load on each contact in the two states of self stress. The distance from any line through the origin thus represents the value of the forces in some linear combination of the two states of self stress. Therefore, we represent linear combinations of $F_1$ and $F_2$ graphically by drawing sloped lines through the origin. The black lines represent the two linear combinations of force eigenvectors at the boundary of the allowed region of force space, shaded green. The two contacts which define these boundaries are shown in red. The green line shows the measured physical forces in the packing represented as a linear combination of the states of self stress. Inset shows a region near the origin in greater detail. For a more pedagogical explanation of this manner of visualizing the FNE, see figure 1 in \cite{sartor_direct_2020}.}
\label{plot:fig1}
\end{figure}







In order to achieve mechanical stability, any system must have at least as many constraints as degrees of freedom. In a granular system, these constraints are borne by the contacts, and for a $d$ dimensional system of frictionless spheres, the minimum number of contacts $N_c^* \sim Nd$ \cite{maxwell_calculation_1864}. Any granular system that posseses more than this minimum number of contacts will have a resultant indeterminacy in its force networks as there must exist multiple linearly independent solutions for force balance. Near jamming, the overlaps (or deformations) between particles are much smaller than the interparticle distances. Due to this separation of scales, the forces in a system can be decoupled from the particle positions, and therefore can be considered to be a random instantiation within the space of the force indeterminancy \cite{snoeijer_force_2004,tighe_force_2010,tighe_stress_2011,sartor_direct_2020}. This is the motivation for the force network ensemble (FNE), which samples all valid force networks in the spring representation of a packing with equal probability.

The rigidity matrix, $\mathcal{R}$, represents a granular system as an unstressed spring network by encoding the normalized contact force vectors $\hat n_{ij}$ between pairs of particles $i$ and $j$ as 
\begin{align}
 \mathcal{R}^{k\gamma}_{\langle ij \rangle} &= \left(\delta_{j}^k-\delta_{i}^k\right)\hat n^\gamma_{ij},
\end{align}
where $k$ indexes particles, $\gamma$ indexes spatial dimensions, and $\delta$ is the Kronecker delta \cite{ellenbroek_rigidity_2015,charbonneau_jamming_2015,f._hagh_broader_2019, sartor_direct_2020}. $\mathcal{R}$ has one column for every degree of freedom and one row for every constraint.  Thus, a granular system consisting of $N$ particles in dimension $d$ with $N_c$ inter-particle contacts will have an $\mathcal{R}$-matrix that is $Nd \times N_c$. In periodic boundary conditions the minimum number of contacts required for stability is
\begin{align}
 N_c^* = Nd - d + 1,
\end{align}
\cite{goodrich_finite-size_2012} and a system with exactly $N_c^*$ contacts will have one stable force network configuration. For each additional contact a system has in excess of $N_c^*$, the associated unstressed spring network will have an additional linearly independent mechanically stable force network. These linearly independent force networks are referred to as the ``states of self stress'' or SSS of the system. These can be easily computed as they are the left singular vectors of $\mathcal{R}$ associated with the zero singular values of $\mathcal{R}$, i.e. the vectors $F_i$ such that $F_i \mathcal{R}   = \vec{0}$. While these SSS in general contain compressional as well as tensional forces, physical packings of frictionless spheres are constrained to compressional forces. Thus we consider the FNE to be the set of linear combinations of the SSS which are positive semidefinite.

In previous work, we demonstrated that by considering the geometric nature of the SSS of a system, one can calculate the volume of the positive semidefinite linear combinations and from that the entropy of the force networks \cite{sartor_direct_2020}. Further contemplation of this geometry has led us to examine the boundaries of this volume, which correspond to sets of contacts which are extraneous to the mechanical stability of the system. In particular, we focus on systems with exactly 2 states of self stress (2SSS), which are thus geometrically confined to have exactly two such boundaries. Each boundary corresponds to a set of contacts (typically each containing exactly one contact) which are unnecessary for mechanical stability of the packing. The breaking of this unnecessary contact results in a packing with just a single SSS (1SSS). In this work, we show that (i) between rearrangements, the force network ensemble of a system is stable under decompression, and (ii) that these boundaries of the volume of allowed force space identify the contacts that may be broken under decompression.








We use pyCudaPacking~\cite{charbonneau_universal_2012}, a GPU-based simulation engine, to generate energy minimized harmonic soft sphere packings. We do so for numbers of particles $N=\{128, 1024, 8192\}$ and in spatial dimension $d=3$. 
The particles are monodisperse and subject to periodic boundary conditions. We minimize the packings using the FIRE minimization algorithm \cite{bitzek_structural_2006} using quad precision floating point numbers in order to achieve sufficient resolution on the contact network near the jamming point.

\begin{figure}[t!]
\includegraphics[width=\columnwidth, trim=94 242 130 260, clip]{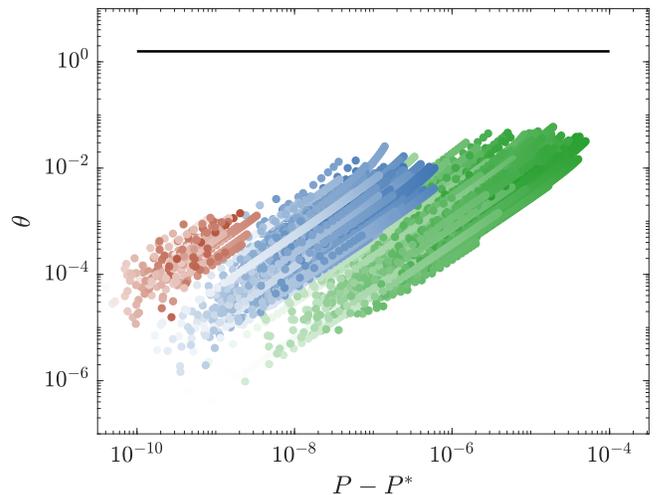}
\caption{Angle $\theta$ between the force spaces of a system at each pressure versus the final force space configuration. At $P=P^*$, $\theta=0$ by construction, but as the pressure increases towards the formation of a new contact, this angle increases. However, it is always very far from perpendicular, which would have a value of $\pi/2$ and is shown as the black line. Colors represent number of particles, with $N=128,1024,8192$ shown as green, blue, and red respectively. The opacity represents the absolute pressure of each, for visual distinguishability.}
\label{plot:anglevpressure}
\end{figure}

Using the same methods as in Refs. \cite{sartor_mean-field_2021,morse_echoes_2017}, we start with randomly distributed initial positions at a packing fraction $\varphi$ far above jamming and apply a search algorithm to create systems approximately logarithmically spaced in excess packing fraction, $\Delta\varphi$. At each step we use the known power law relationship between energy and $\Delta\varphi$ to calculate an estimate of the jamming packing fraction $\varphi_j$. We use this estimate to approximate $\Delta\varphi$ and determine the next value of the $\varphi$ in an effort to logarithmically space $\Delta\varphi$ values, with a very fine spacing in $\Delta\varphi$ (100 steps/decade) so that we may probe the dynamics of the transition from 2SSS to 1SSS. We then adjust the packing fraction to this value of $\varphi$ by uniformly scaling particle radii and minimizing the system. We continue this process until the system has exactly one state of self stress. We generate datasets of 500 systems at $N={128, 1024}$ and 100 systems at $N=8192$.  We measure the pressure $P$ from the trace of the stress tensor as in Ref. \cite{ohern_jamming_2003} and denote the pressure at which the system transitions from 2SSS to 1SSS as $P^*$.


The force network ensemble is the set of linear combinations of these SSS for which all forces are positive semi-definite. This defines a region in force space, the boundaries of which are the linear combinations of the SSS that bring the load on a contact or set of contacts to zero. 
In a system with 2SSS, one can exploit the orthogonality of the SSS to choose the linear combination that yields zero force on any given contact. The imposition of this constraint necessarily reduces the number of SSS by one, and within the context of the force network ensemble is equivalent to breaking a contact. However, with most contacts, this will result in negative forces (i.e. tensile loads) on some of the contacts unless the contact chosen is on the boundary of the allowed volume of force space. We show this geometrically in figure \ref{plot:fig1}.

As a system decompresses, the geometry of the contact network changes. This is reflected in the rigidity matrix and thus results in changes in the null space of the loads. Since our systems are at 2SSS, this null space is always a two-dimensional plane within the $N_c$ dimensional space of allowed loads on bonds. We characterize how this plane evolves by computing its angle $\theta$ relative to its final configuration at $P=P^*$. We define the angle $\theta$ between 2SSS force spaces $\mathcal{F}$ and $\mathcal{G}$ (with basis vectors $\hat F_i$ and $\hat G_i$) as~\cite{jordan_essai_1875}:
\begin{align}
\cos (\theta) = \frac{|\wedge^2 g(\mathcal{F}, \mathcal{G})|}{\sqrt{ \wedge^2 g (\mathcal{F},\mathcal{F}) \wedge^2 g(\mathcal{G},\mathcal{G})}}
\end{align}
where
\begin{align}
(\wedge^2 g) (\mathcal{F}, \mathcal{G}) = \det\begin{pmatrix} \hat F_1 \cdot \hat G_1 & \hat F_1 \cdot \hat G_2 \\ \hat F_2 \cdot \hat G_1 & \hat F_2 \cdot \hat G_2 \end{pmatrix}.
\end{align}
While uncorrelated 2 dimensional planes drawn through the $N_c$ dimensional force space are nearly perpendicular (i.e. $\theta \sim \frac{\pi}{2}$), we show in figure \ref{plot:anglevpressure} that the force space volumes of our systems are always nearly aligned, even over wide ranges of pressure. This shows that the force network ensemble of a packing is stable under decompression, at least between rearrangements. Thus, it should be possible to use the force network ensemble to predict the evolution of the physical forces as the pressure is varied.

\begin{figure}[h!]
\includegraphics[width=\columnwidth, trim=98 140 139 101, clip]{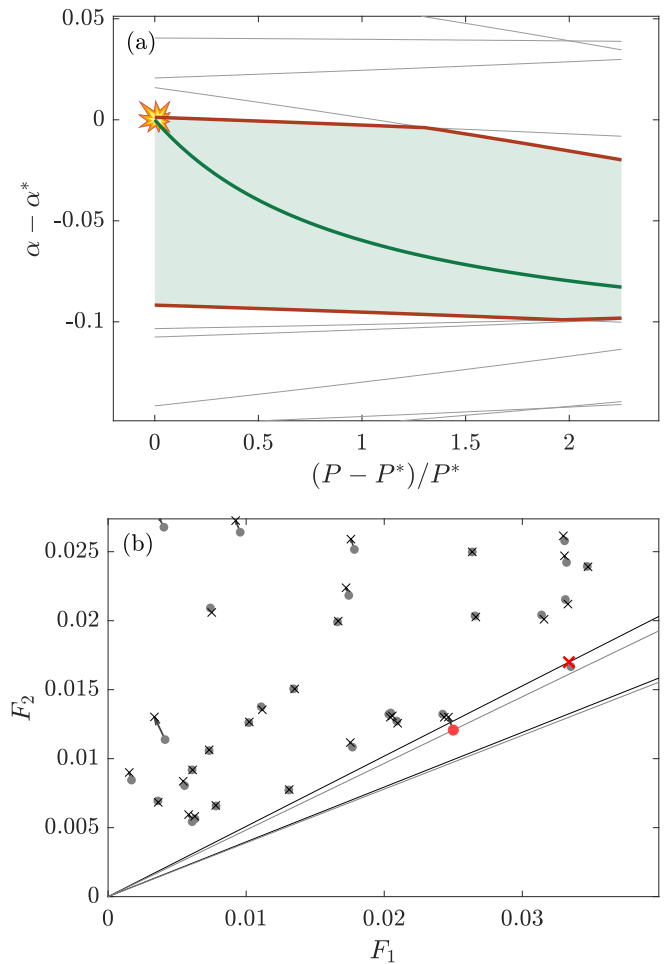}
\caption{a) Evolution of force space under decompression for the system in figure \ref{plot:fig1} over the range of pressures for which the system is at 2SSS. Position in force space is shown as mixing angles $\alpha-\alpha^*$ where $\alpha^*$ is the mixing angle of the system at the contact breaking event. The mixing angle for the physical forces of a system is shown in green and the green shaded region shows all mixing angles which correspond to positive definite force networks. The boundaries of this allowed force space correspond to positive semi-definite force networks and are shown in red. Mixing angles which would bring other contacts to zero resulting in some contacts having negative force are shown in grey. Each contact in the system has such a path, but most exist outside the shown range of mixing angles. b) The evolution of the system in force space, shown zoomed in for more detail as in the inset to figure \ref{plot:fig1}. The highest and lowest pressures at which the system is at 2SSS are shown as grey circles and black x's respectively, with arrows between them. The ``breakable contacts'' on the edge of force space are shown in red. While no force point has moved very far, the breakable contact at the boundary is seen to have been exchanged. We note that as the SSS space is degenerate, there is a free overall rotation in each system. We thus choose $F_1$ and $F_2$ at each pressure to minimize the motion of the force points.}
\label{plot:trajectory}
\end{figure}

In a system at 2SSS, the space of normalized linear combinations of the SSS is a one dimensional space of rotations, as any stable force configuration $f$ can be described by a mixing angle $\alpha$ such that $f = \sin(\alpha) F_1 + \cos(\alpha) F_2$, where $F_1$ and $F_2$ are the linearly independent SSS. While the physical forces in the packing are instead calculated from the overlaps, they represent a stable force network and as such we are always able to express them with a mixing angle in this way, up to machine precision and an overall scale factor. In the intervals between contact changes, we may thus consider the physical forces in the packing as flowing within this space of SSS, which is only gently changing as shown in figure \ref{plot:anglevpressure}. In figure \ref{plot:trajectory}(a), we show the mixing angles that describe the position within this space for the physical forces of a 2SSS system as it decompresses towards $P=P^*$. By following these mixing angles as a system decompresses, we see how the physical forces in the system approach and reach one of the 1SSS states on the boundaries of the 2SSS space. As an example, we can follow the upper red curve in figure \ref{plot:trajectory}(a) down in pressure towards the contact break, and we see that this system has a kink around $(P-P^*)/P^* \sim 1.3$. This arises from the exchange of the contact that originally formed the boundary of the allowed force space for another, which can be seen graphically in figure \ref{plot:trajectory}(b). Thus a prediction made with the boundary contact above that pressure will fail. The interchange in this manner of boundary contacts with other contacts that were intitially near the boundary is relatively unusual and is the sole failure mode of our prediction.

\begin{figure}[t!]
\includegraphics[width=\columnwidth, trim=104 240 123 260, clip]{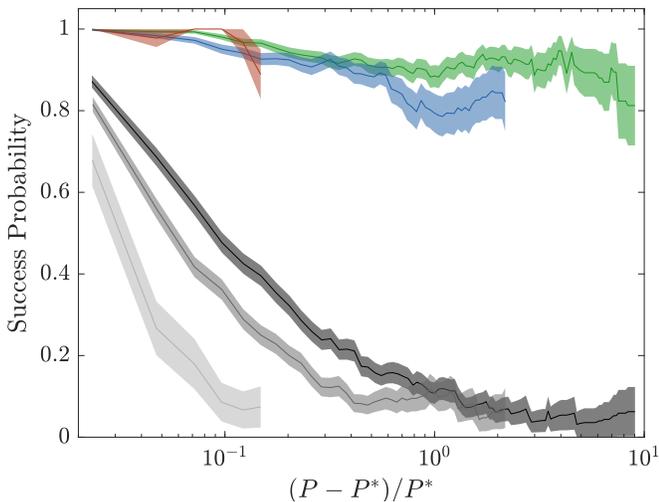}
\caption{Probability of predicting contact breaking by FNE shown against scaled pressure, for systems decompressing from 2SSS to 1SSS. Colors represent number of particles as in fig. \ref{plot:anglevpressure}, with green, blue, and red representing $N=128,1024,8192$ respectively. Probability of smallest force in the system breaking is shown in dark grey for $N=128$, medium grey for $N=1024$, and light grey for $N=8192$. The FNE prediction is accurate even at high pressures, while the smallest contact approach quickly loses predictive power. Note that because systems above 2SSS are not included, there are fewer systems at higher pressures.}
\label{plot:predictionProb}
\end{figure}

In a 2SSS system, there are always exactly two boundaries on the edge of the force space which correspond to two sets of ``breakable contacts.'' Each of these sets of contacts usually contains just one contact, but sometimes breaking a contact will form a rattler particle, all of whose contacts will thus be in the set of breakable contacts. In figure \ref{plot:predictionProb}, we show the probability that one of these sets of ``breakable contacts'' is in fact broken under decompression from 2SSS to 1SSS. We find that the contacts predicted by the FNE are strongly predictive (greater than $80\%$) over the full range of pressures for which the system has 2SSS. This is in sharp contrast to the na\"ive prediction from affine response, that the smallest contact will break. Affine response is predictive very close to the contact breaking event, but falls to zero at higher pressures.

We examined the real space correlations between pairs of breakable contacts and found no correlation in position or angle subtended between contact vectors. However, we find that breakable contacts are more likely to occur between particles with higher than average number of contacts: while a typical particle (in 3 dimensions) has an average contact number of 6 (i.e. $2d$), a particle with a breakable contact has an average number of contacts of $\sim 6.4$. This shows that particles with an above average number of contacts are more likely to have a contact that is unnecessary for system stability. This stands in contrast to the soft spot literature, in which particles with fewer contacts are identified as more likely to rearrange \cite{manning_vibrational_2011,ridout_correlation_2020,rocks_learning-based_2021,richard_predicting_2020}. However, this is to be expected, since soft spots exclusively identify instabilities, whereas the contact changes predicted by the FNE primarily comprise stable transitions to configurations with fewer contacts rather than instabilities.

\begin{figure}[t!]
\includegraphics[width=\columnwidth, trim=103 239 130 252, clip]{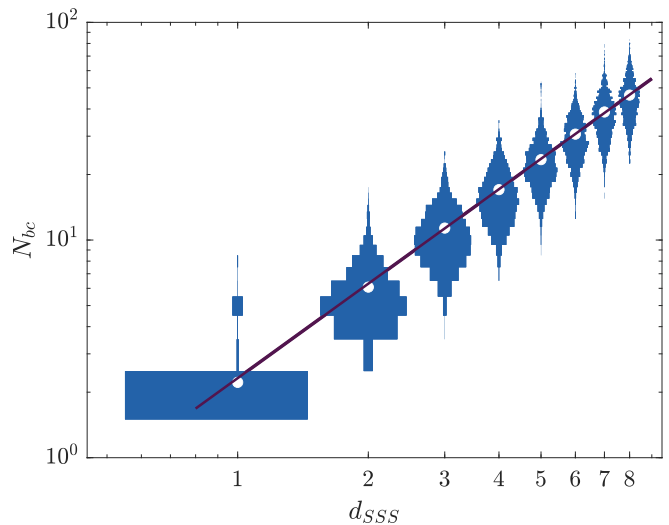}
\caption{Histogram of the number of breakable contacts, $N_{bc}$, at each number of dimensions of the normalized space of states of self stress, $d_{SSS}$, for $N=128$. We see that when this space is 1 dimensional (i.e. 2SSS) we usually have two breakable contacts on the boundary. As this space increases in dimension, however, many more contacts quickly become breakable. The purple line shows the empirical fit:  $N_{bc} \approx 2.43d^{1.44}$.}
\label{plot:numBreakable}
\end{figure}

We have shown that for a system with 2SSS, the two breakable contacts at the boundaries of the force space are excellent predictors of contact breaking events. For systems at higher pressure, however, there are more contacts and thus a larger number of SSS. In figure \ref{plot:numBreakable}, we examine the relationship between the number of predicted breakable contacts, $N_{bc}$ and the number of states of self stress, $N_{SSS}$. 
Because the dimensionality of the space of normalized states of SSS is one less than $N_{SSS}$, we examine the scaling of $N_{bc}$ with this dimension, $d_{SSS} = N_{SSS}-1$. 
At 2SSS, the breakable contacts can be understood as existing at the endpoints of a line which represents the space of allowed mixing angles. This typically results in two breakable contacts (i.e. one at each endpoint), but also may result in $N_{bc}\geq 5$ by creating a rattler at one endpoint (i.e. four or more contacts lost at that endpoint). We rarely find $N_{bc}=3,4$ which can only arise from degeneracies or numerical instabilities in the 1SSS force network.
At 3SSS, the boundaries of allowed force space can be considered as the edges of a polygon, and at higher SSS a polytope. These must have at least $d_{SSS}$ boundaries, but may have arbitrarily many and thus arbitrarily many breakable contacts.
We characterize the distribution of number of breakable contacts at each SSS by their mean, and find the means to be remarkably well fit by $N_{bc} \approx  2.32(d_{SSS})^{1.44}$. 

We conjecture that even at higher than 2SSS, the FNE may be used to predict contact breaking events.
It has recently been shown that $86\%$ of contact breaking events are reversible network events rather than rearrangements \cite{morse_differences_2020,tuckman_contact_2020}. We would thus expect that these reversible contact breaking events are well described by the FNE, and as such one could use our methods to predict the possible final 1SSS systems from a system with several contacts over 2SSS, with a success probability that scales as $P \sim 0.86 ^{d_{SSS}}$. We note however that while the physical forces of our packings were found precisely in the force network ensemble, this may fail at significantly higher pressures, because the force network ensemble is calculated from the unstressed spring network representation of a packing, whereas there exists a prestress on the physical forces.

We have shown that the force network ensemble of jammed systems remains approximately static between contact change events, and as such may be used to identify defective contacts. We have further shown that for 2SSS systems, these identified defective contacts are highly likely to break under decompression. While in spirit these defects can be thought of as analogous to soft spots, we emphasize two key differences here: (1) Soft spots identify locations of instabilities, while the force network ensemble identifies locations of both instabilities and stable contact network changes, and (2) While soft spots may be identified using just local information, the FNE approach by definition invokes global information. One might wish to use local information to find these defects, and there exists local structure to the force networks when the system is far from jamming \cite{sussman_spatial_2016}. However, at the jamming point the system becomes marginal, and as such any change in the contact network impacts the whole force network. For this reason we believe that it would not be possible to identify these defective contacts near jamming without global information. 

By restricting our study to 2SSS systems under decompression, we have explored only a limiting case of the space of the force network ensemble. 
We hope however that this serves as a gateway for future work, especially in exploring the force network landscape for systems with a greater number of SSS, where many more defects are predicted. One could also apply a modified version of these methods to identify force network defects in systems under shear.
Another interesting and unexplored avenue is the search for force network defects in ultrastable systems \cite{kapteijns_fast_2019,hagh_transient_2021,ninarello_models_2017}, which have more contacts near jamming than random packings, and as such the defects may be delocalized or poorly defined. 
Our protocol identifies both stable contact changes and those that lead to stabilities. One further interesting remaining question is whether these can be differentiated from each other via the force network ensemble, giving a more complete picture of the defects within the system.

We thank Varda Hagh, Sean Ridout, Brian Tighe, Rafael D\'iaz Hern\'andez Rojas, and Peter Morse for valuable discussion and feedback. This work is supported by the Simons Collaboration on Cracking the Glass Problem via award No. 454939.


\bibliography{forceSpacePath}

\end{document}